\documentclass[11pt,twoside]{article}

\usepackage{asp2014}

\aspSuppressVolSlug
\resetcounters

\begin{document}

\title{\texttt{Spherinator} and \texttt{HiPSter}: Representation Learning for Unbiased Knowledge Discovery from Simulations}

% Note the position of the comma between the author name and the 
% affiliation number.
% Authors surnames should come after first names or initials, eg John Smith, or J. Smith.
% Author names should be separated by commas.
% The final author should be preceded by "and".
% Affiliations should not be repeated across multiple \affil commands. If several
% authors share an affiliation this should be in a single \affil which can then
% be referenced for several author names. If only one affiliation, no footnotes are needed.
% See ManuscriptInstructions.pdf and ASP's manual2010.pdf 3.1.4 for more details
\author{Kai L. Polsterer, Bernd Doser, Andreas Fehlner and Sebastian Trujillo-Gomez}
\affil{Heidelberg Institute for Theoretical Studies, Heidelberg, Baden-W\"urttemberg, Germany; \email{kai.polsterer@h-its.org}}

% This section is for ADS Processing.  There must be one line per author. paperauthor has 9 arguments.
\paperauthor{Kai Lars Polsterer}{kai.polsterer@h-its.org}{https://orcid.org/0000-0002-3435-1912}{HITS gGmbH}{Astroinformatics}{Heidelberg}{BW}{69118}{Germany}
\paperauthor{Bernd Doser}{bernd.doser@h-its.org}{0000-0002-3443-5913}{HITS gGmbH}{Astroinformatics}{Heidelberg}{BW}{69118}{Germany}
\paperauthor{Andreas Fehlner}{andreas.fehlner@h-its.org}{0000-0003-0448-7138}{HITS gGmbH}{Astroinformatics}{Heidelberg}{BW}{69118}{Germany}
\paperauthor{Sebastian Trujillo-Gomez}{sebastian.trujillogomez@h-its.org}{0000-0003-2482-0049}{HITS gGmbH}{Astroinformatics}{Heidelberg}{BW}{69118}{Germany}

% There should be one \aindex line (commented out) for each author. These are used to
% build up the author index for the Proceedings. The surname must come first, followed by
% initials. Note the use of ~ before each initial to control spacing.
% The \author entries (see above) have surname last. These \aindex entries have
% surname first.
% The Aindex.py command willl create them for you after you have constructed the \author
% The first entry should be the first author, for bold-facing the author index page-reference

%\aindex{FistAuthor1,~S.~A.}
%\aindex{Author2,~S.~B.}
%\aindex{Author3,~S.}

\begin{abstract}
Simulations are the best approximation to experimental laboratories in astrophysics and cosmology.
However, the complexity, richness, and large size of their outputs severely limit the interpretability of their predictions.
We describe a new, unbiased, and machine learning based approach to obtaining useful scientific insights from a broad range of simulations.
The method can be used on today’s largest simulations and will be essential to solve the extreme data exploration and analysis challenges posed by the Exascale era.
Furthermore, this concept is so flexible, that it will also enable explorative access to observed data.

Our concept is based on applying nonlinear dimensionality reduction to learn compact representations of the data in a low-dimensional space.
The simulation data is projected onto this space for interactive inspection, visual interpretation, sample selection, and local analysis.
We present a prototype using a rotational invariant hyperspherical variational convolutional autoencoder, utilizing a power distribution in the latent space, and trained on galaxies from IllustrisTNG  simulation.
Thereby, we obtain a natural \emph{Hubble tuning fork} like similarity space that can be visualized interactively on the surface of a sphere by exploiting the power of HiPS tilings in Aladin Lite.
\end{abstract}

\section{Introduction}

Cosmological hydrodynamical simulations, or simulations in general, are excellent numerical laboratories to understand astrophysical processes like the formation of galaxies or the large-scale structure.
In accordance with Richard Feynman's quote "What I cannot create, I do not understand", those virtual laboratories are essential to understand what we observe.
Those simulations provide access to a highly detailed realization of simulated universes across a vast range of spatial and temporal scales, from shortly after the Big Bang until the present day, containing all main matter components, e.g. dark matter, gas, stars, and black holes.
Each of the simulation time-steps is extremely information-rich, including 6D phase space positions and velocities, as well as dozens of physical properties for all components, e.g. density, temperature, metallicity, individual element abundances, etc.). 

State-of-the-art large-volume cosmological simulations currently model the evolution of the Universe using more than $10^{11}$ particles and produce petabytes of data.
This level of data size and complexity already exceeds by far the exploration, synthesis, and interpretation capacity of humans; with the rise of the Exascale era rendering traditional analysis techniques obsolete \citep{Nelson2019}.
Moreover, architectures and I/O formats vary greatly across simulation codes, imposing a barrier to the application of code-specific tools more generally to all cosmological datasets.
Simulation data is typically represented in a very compressed format in catalogs, often representing the rich multidimensional data by single scalar representations (e.g. galaxy and DM halo properties).
By utilizing dimensionality reduction techniques, we strive for automatically learning a more meaningful compression and embedding of the original data.
Machine learning (ML) provides a set of powerful tools to achieve such a compression to an arbitrary number of dimensions, while ensuring that similar objects have a similar compressed representation.
This is comparable to empirical found relations w.r.t. galaxy properties, like the Hubble diagram, the galaxy main sequence, the mass-metallicity relation, the Tully-Fisher relation, or the fundamental plane.

As many tools in astronomy exist to visualize and handle data on a sphere, we decided to evaluate the use of HiPS \citep{2015A&A...578A.114F} to represent an ML derived projection on a spherical surface.
Especially, the hierarchical nature of HiPS allows to iteratively refine the view and thereby provide explorative access to the complexly structured data.
This is an essential contribution to the aspects of findability and accessibility as part of the FAIR data principles \citep{wilkinson2016fair}.
The dimensionality reduction is handled by \texttt{Spherinator} while the HiPS tile representation is created by \texttt{HiPSter}.

\section{\texttt{Spherinator}}

% Structure:
% - Variational autoencoder
%   - Compare with non-variational AE 
% - (hyper-)spherical latent space
% - MSE reconstruction loss
% - Kullback Leibler divergence
% - Balancing KL and MSE loss
%   - posterior collapse
%   - KL annealing
%   - beta/lambda factor
% - Convolutional layer
% - Replace KL by direct use of z_scale / kappa

To achieve a meaningful embedding in a lower dimensional space, neural networks have proven to be powerful ML approaches.
In recent years, hyperspherical variational autoencoders (HVAE) have emerged as a powerful tool in the field of generative modeling \citep{DeCao2020, Davidson2018}, utilizing a spherical surface for the embedding.
The hyperspherical latent space enables more efficient and meaningful exploration of the latent representation, as the hyperspherical space is continuous without boundaries like the traditional Euclidean space but in contrast is constrained in size.
The presented first version of \texttt{Spherinator} uses PyTorch Lightning to implement a convolutional neural network (CNN) based variational autoencoder (VAE) with a spherical latent space.

We implemented an encoder with a five layer cascade of convolutional and pooling layers [128, 64, 32, 16, 8, 4] and a corresponding decoder.
The bottleneck was designed to represent a density distribution on a spherical surface using a simple renormalized 3D vector of unity length for the coordinates $\mu$ and a rotationally symmetric concentration parameter $\kappa$.
To achieve rotational invariance, the reconstruction loss for a set of rotated images and their reconstruction was calculated, and only the version with the lowest loss was used for optimizing the weights and biases of the model.
The loss function consists of two parts: the reconstruction loss and the Kullback-Leibler (KL) divergence.

\begin{displaymath}
L = L_{recon} + \lambda \cdot L_{KL}
\end{displaymath}

\noindent
The reconstruction loss is measured using the pixel wise root-mean-square deviation (RMSE).
This loss and the KL divergence have to be balanced, which is achieved via the factor $\lambda$, which helps to ensure a meaningful latent space \citep{Asperti2020}.
The KL divergence measures the difference between two probability distributions.
A Gaussian reference distribution is used for Euclidean latent spaces, while a uniform reference distribution is taken for hyperspherical latent spaces.

\noindent
The power-spherical distribution is defined as 

\begin{displaymath}
p_{X}(x; \mu, \kappa) = N_{X}(\kappa, d)^{-1}(1 + \mu^{\top}x)^{\kappa}
\end{displaymath}

\noindent
with direction $\mu$ and concentration $\kappa$ and normalization factor $N_{X}$.
Since the KL divergence between a uniform distribution and a power spherical distribution is independent of the direction $\mu$, the regularization term can be expressed directly using $\kappa$.
We used the Adam optimizer with a learning rate scheduler from $10^{-3}$ to $10^{-5}$ and batch size 256 to train the model for 200 epochs.
Hereby, we calculated 36 rotated versions of every input dataset, creating versions for every $10\deg$ of rotation.
We found $\lambda=10^{-3}$ to be a good choice to produce a reasonable embedding as depicted on our poster.

\articlefiguretwo{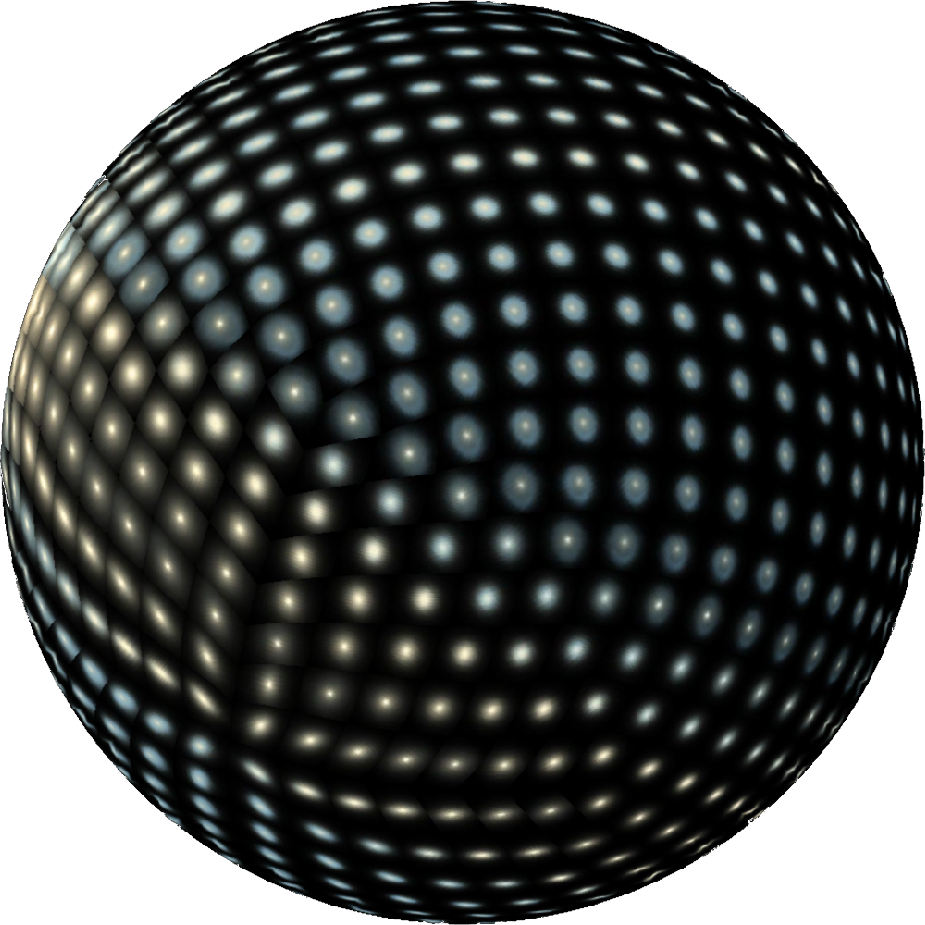}{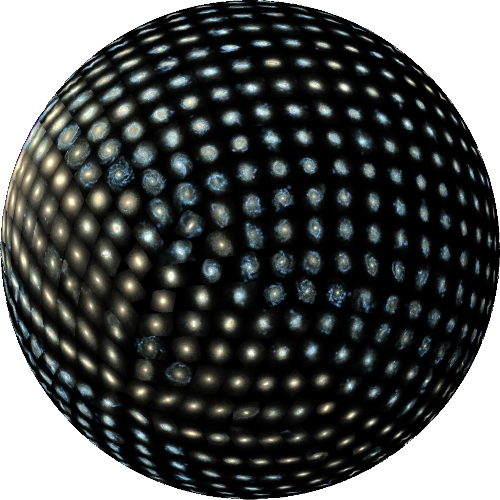}{P404_fig1}{ Output of \texttt{HiPSter}. \emph{Left:} Reconstruction of the trained HVAE model. \emph{Right:} Embedded original images of the galaxies closest to the center of each tile.}

\section{\texttt{HiPSter}}

As soon as a model is trained, the \texttt{HiPSter} is used to create an explorable representation of the projected data-set.
Hereby, we utilize the HiPS standard to store the results in a hierarchical form.
This enables us to represent extremely large data-sets in a very compact way, a first step to the data we expect from Exascale simulations.
To generate all necessary files, the following tasks had been implemented.

\noindent
\textbf{Model Representation:} Based on the trained autoencoder, we utilize only the generative decoder part to produce an image for the center coordinates of each healpix cell.
Those images are stored following the HiPS standard alongside with a properties file.

\noindent
\textbf{Catalog Creation:} The encoder part is used to embed all input images on the sphere.
This task is done, using the same technique for rotation invariance as described for the training of the model.
The resulting coordinates from the bottleneck as well as some meta-information are written to a CSV-file and a VO-table.

\noindent
\textbf{Image Embedding:} Based on the created catalog file, for each order of the HiPS, the tiles are filled with the input image that projects closest to the center of the corresponding healpix cell.
In case no input image maps to a cell, the corresponding tile is left empty.
This generates the representation shown in Figure~\ref{P404_fig1}.

\noindent
\textbf{Thumbnails \& Originals:} We generate JPEG images for a hover on presentation and to allow inspecting the original uncropped and unscaled input images within AladinLite.

\noindent
\textbf{Allsky Images:} To support Aladin Sky Atlas on the desktop, overview images are produced for every order following the HiPS standard.

\noindent
\textbf{Deployment:} To deploy the produced two HiPS tilings as well as the catalog files, an HTML-file is automatically created.
This file utilizes AladinLite to provide direct visual explorative access to the representation learned with the \texttt{Spherinator} as presented on \url{https:\\space.h-its.org}

\section{Conclusions}
When the tiles and catalogs are access via Aladin Sky Atlas on the desktop, users can annotate regions of interest, create own catalogs by using multi-order coverage maps, search by similarity by querying sources in the vicinity through e.g. a cone search, or identify outliers by looking at those sources with the largest reconstruction errors.
A whole new data access workflow arises from the representation of morphological structures as coordinates on a sphere, enabling the realization of very diverse scientific use-cases.

\acknowledgements 

We gratefully acknowledge the generous and invaluable support of the Klaus Tschira Foundation.
This work has received funding from the European High Performance Computing Joint Undertaking (JU) and Belgium, Czech Republic, France, Germany, Greece, Italy, Norway, and Spain under grant agreement No101093441.
Views and opinions expressed are however those of the author(s) only and do not necessarily reflect those of the European Union or the European High Performance Computing Joint Undertaking (JU) and Belgium, Czech Republic, France, Germany, Greece, Italy, Norway, and Spain.
Neither the European Union nor the granting authority can be held responsible for them.
This research has made use of "Aladin sky atlas" developed at CDS, Strasbourg Observatory, France.
We are grateful to Thomas Boch for help with AladinLite, and to Dylan Nelson and the TNG Collaboration for providing access to their data and visualizations.
Code of \texttt{Spherinator} and \texttt{HiPSter} are available at \url{https://github.com/HITS-AIN/Spherinator}

\bibliography{P404}  % For BibTex

\begin{thebibliography}{}
\expandafter\ifx\csname natexlab\endcsname\relax\def\natexlab#1{#1}\fi
\expandafter\ifx\csname url\endcsname\relax
  \def\url#1{\texttt{#1}}\fi
\expandafter\ifx\csname urlprefix\endcsname\relax\def\urlprefix{URL }\fi
\providecommand{\eprint}[2][]{\url{#2}}

\bibitem[{Asperti \& Trentin(2020)}]{Asperti2020}
Asperti, A., \& Trentin, M. 2020, IEEE Access, 8, 199440. \urlprefix\url{http://arxiv.org/abs/2002.07514}

\bibitem[{Cao \& Aziz(2020)}]{DeCao2020}
Cao, N.~D., \& Aziz, W. 2020, Proceedings of the 37th International Conference on Machine Learning, INNF+. \urlprefix\url{http://arxiv.org/abs/2006.04437}

\bibitem[{Davidson et~al.(2018)Davidson, Falorsi, Cao, Kipf, \& Tomczak}]{Davidson2018}
Davidson, T.~R., Falorsi, L., Cao, N.~D., Kipf, T., \& Tomczak, J.~M. 2018, 34th Conference on Uncertainty in Artificial Intelligence (UAI-18). \urlprefix\url{http://arxiv.org/abs/1804.00891}

\bibitem[{{Fernique} et~al.(2015){Fernique}, {Allen}, {Boch}, {Oberto}, {Pineau}, {Durand}, {Bot}, {Cambr{\'e}sy}, {Derriere}, {Genova}, \& {Bonnarel}}]{2015A&A...578A.114F}
{Fernique}, P., {Allen}, M.~G., {Boch}, T., {Oberto}, A., {Pineau}, F.~X., {Durand}, D., {Bot}, C., {Cambr{\'e}sy}, L., {Derriere}, S., {Genova}, F., \& {Bonnarel}, F. 2015, \aap, 578, A114. \eprint{1505.02291}

\bibitem[{{Nelson} et~al.(2019){Nelson}, {Pillepich}, {Springel}, {Pakmor}, {Weinberger}, {Genel}, {Torrey}, {Vogelsberger}, {Marinacci}, \& {Hernquist}}]{Nelson2019}
{Nelson}, D., {Pillepich}, A., {Springel}, V., {Pakmor}, R., {Weinberger}, R., {Genel}, S., {Torrey}, P., {Vogelsberger}, M., {Marinacci}, F., \& {Hernquist}, L. 2019, \mnras, 490, 3234. \eprint{1902.05554}

\bibitem[{Wilkinson et~al.(2016)Wilkinson, Dumontier, Aalbersberg, Appleton, Axton, Baak, Blomberg, Boiten, da~Silva~Santos, Bourne et~al.}]{wilkinson2016fair}
Wilkinson, M.~D., Dumontier, M., Aalbersberg, I.~J., Appleton, G., Axton, M., Baak, A., Blomberg, N., Boiten, J.-W., da~Silva~Santos, L.~B., Bourne, P.~E., et~al. 2016, Scientific data, 3

\end{thebibliography}

% if we have space left, we might add a conference photograph here. Leave commented for now.
% \bookpartphoto[width=1.0\textwidth]{foobar.eps}{FooBar Photo (Photo: Any Photographer)}

\end{document}